\documentclass[journal]{IEEEtran}
\usepackage{cite}

\newcommand{\tnr}[1]{{\textnormal{#1}}}

\usepackage{tikz,pgfplots}
\usepgfplotslibrary{units}
\usetikzlibrary{positioning,shadows,shapes,fit,arrows,backgrounds}

\usetikzlibrary{external}
\colorlet{figyellow}{yellow!40!white}
\colorlet{figred}{red!40!white}
\colorlet{figblue}{blue!20!white}
\colorlet{figgreen}{green!30!white}
\colorlet{figgray}{black!10!white}

\ifCLASSINFOpdf
\else
\fi

\usepackage{amsmath}
\usepackage[caption=false]{subfig}

\def\docTitle{The\ Benefit\ of\ Split\ Nonlinearity\ Compensation\ for\ Optical\ Fiber\ Communications}

\begin{document}
%
\title{\docTitle}
%
%
%

\author{Domani\c{c}~Lavery,~\IEEEmembership{Member,~IEEE,}
        David~Ives,
	Gabriele~Liga,~\IEEEmembership{Student~Member,~IEEE,}
        Alex~Alvarado,~\IEEEmembership{Senior~Member,~IEEE,}
        Seb~J.~Savory,~\IEEEmembership{Senior~Member,~IEEE,}
        and~Polina~Bayvel,~\IEEEmembership{Fellow,~IEEE}
\thanks{This work was supported by the U.K. Engineering and Physical Sciences Research Council (EPSRC) under grants EP/J017582/1 (UNLOC) and {EP/L026155/1} (INSIGHT).}
\thanks{The authors are with the Department of Electronic and Electrical
Engineering, Optical Networks Group, University College London, London.
WC1E 7JE, U.K. (e-mail: d.lavery@ee.ucl.ac.uk; alex.alvarado@ieee.org; {david.ives.10, gabriele.liga.11, s.savory, p.bayvel}@ucl.ac.uk).}
}

\maketitle

\begin{abstract}
In this Letter we analyze the benefit of digital compensation of fiber nonlinearity, where the digital signal processing is divided between the transmitter and receiver.  The application of the Gaussian noise model indicates that, where there are two or more spans, it is always beneficial to split the nonlinearity compensation. The theory is verified via numerical simulations, investigating transmission of single channel 50~GBd polarization division multiplexed 256-\textit{ary} quadrature amplitude modulation over 100~km standard single mode fiber spans, using lumped amplification.  For this case, the additional increase in mutual information achieved over transmitter- or receiver-side nonlinearity compensation is approximately 1~bit for distances greater than 2000~km.  Further, it is shown, theoretically, that the SNR gain for long distances and high bandwidth transmission is 1.5~dB versus transmitter- or receiver-based nonlinearity compensation.
\end{abstract}


%
\IEEEpeerreviewmaketitle

\section{Introduction}
%
%
%
%
\IEEEPARstart{R}{ecent} efforts in overcoming the limit to optical communications imposed by fiber nonlinearity can be broadly grouped into two areas: optical and digital techniques.  Optical techniques include, for example, optical phase conjugation (OPC) using twin waves \cite{Liu2013} or OPC devices placed mid-span \cite{Ellis2015_OPC}. Digital techniques include transmitter- \cite{Temprana2015, Roberts2006} and receiver-side \cite{Maher2015} digital nonlinearity compensation (NLC), simple nonlinear phase shifts \cite{Lowery2007, Lavery2015}, perturbation-based precompensation \cite{Gao2013}, {adaptive filtering \cite{Secondini2014a}} and optimum detection \cite{Liga2015}.  With the exception of optimum detection (a special case of receiver-side NLC for single span transmission) the digital signal processing (DSP) techniques are algorithms which invert the propagation equations for the optical fiber, either exactly or with simplifying approximations.

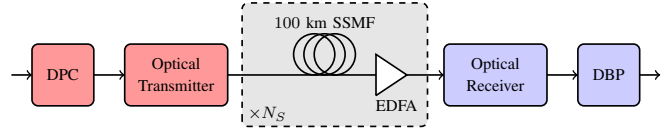
\begin{figure}
\centering
\resizebox{0.79\columnwidth}{!}{%
\centerline{
\footnotesize{
\begin{tikzpicture}[plain/.style={align=center,execute at begin node=\setlength{\baselineskip}{2.5ex}}]
\draw[thick,->] (-175pt,0pt) -- (-165pt,0pt);
\draw[thick,fill=figred,rounded corners=3pt] (-165pt,-15pt) rectangle (-135pt,15pt);
\node[plain] at (-150pt,0pt) {DPC};
\draw[thick,->] (-135pt,0pt) -- (-120pt,0pt);
\draw[thick,fill=figred,rounded corners=3pt] (-120pt,-15pt) rectangle (-70pt,15pt);
\node[plain] at (-95pt,5pt) {Optical};
\node[plain] at (-95pt,-5pt) {Transmitter};
\draw[thick,fill=figgray,dashed,rounded corners=3pt] (-63pt,-25pt) rectangle (27pt,35pt);
\node[plain,anchor=south west] at (-63pt,-25pt) {$\times N_S$};
\node[plain,anchor=south] at (-23pt,20pt) {$100$~km SSMF};
\draw[thick,-] (-28pt,10pt) circle (10pt);
\draw[thick,-] (-23pt,10pt) circle (10pt);
\draw[thick,-] (-18pt,10pt) circle (10pt);
\draw[thick,->] (-70pt,0pt) -- (35pt,0pt); 
\draw[thick,-,fill=white] (2pt,-10pt) -- (17pt,0pt) -- (2pt,10pt) -- (2pt,-10pt);
\node[plain] at (12pt,-15pt) {EDFA};
\draw[thick,fill=figblue,rounded corners=3pt] (35pt,-15pt) rectangle (85pt,15pt);
\node[plain] at (60pt,5pt) {Optical};
\node[plain] at (60pt,-5pt) {Receiver};
\draw[thick,->] (85pt,0pt) -- (100pt,0pt);
\draw[thick,fill=figblue,rounded corners=3pt] (100pt,-15pt) rectangle (130pt,15pt);
\node[plain] at (115pt,0pt) {DBP};
\draw[thick,->] (130pt,0pt) -- (140pt,0pt);
\end{tikzpicture}}}
}
\caption{Transmission model used for investigating the performance of fiber nonlinearity compensation, where the digital nonlinearity compensation is divided between transmitter and receiver.}
\label{fig:config}
\end{figure}

{Consider the model in Fig.~\ref{fig:config}, which shows a transmission link with digital NLC at both the transmitter and the receiver}. To date, the best performing experimentally demonstrated digital technique for receiver-side digital NLC is the digital backpropagation (DBP) algorithm. This algorithm numerically solves the inverse of the optical fiber propagation equations to compensate the linear and nonlinear impairments introduced by the optical fiber transmission; albeit not taking account of amplifier noise ({cf. zero-forcing equalization}).  The demonstrations of multi-channel NLC via receiver-side DBP \cite{Maher2015} which takes account of the inter-channel nonlinear distortions, have recently been repeated using similar digital techniques to predistort for nonlinearity at the transmitter -- digital precompensation (DPC) \cite{Temprana2015}.  We note that, as might be expected due to the symmetry of the transmission link\footnote{As noted in \cite{Lavery2015}, the first span is an exception to link symmetry in that, under the simplifying assumptions of a noiseless transmitter, no polarization mode dispersion and no photon-phonon interactions, the nonlinear interference in this span can be compensated exactly with DPC.}, the experimentally demonstrated performance of the pre- and post-compensation algorithms is similar; achieving a 100\% increase in transmission reach\cite{Temprana2015,Maher2015}.

Whether applying DBP or DPC, additive noise from in-line optical amplifiers is enhanced, which limits the performance of the NLC.  One can, therefore, make a heuristic argument for dividing the DBP equally between transmitter and receiver, thus limiting the noise enhancement in the compensated waveform to the signal-noise interaction present at the center (rather than the end) of the transmission link.

Although split NLC (dividing digital NLC between transmitter and receiver) has previously been considered, experimental implementation was confined to the special case of simplified DSP (single nonlinear phase shift \cite{Lowery2007}) and theoretical analysis considered only residual NLC after OPC \cite{Ellis2015_OPC}. In this Letter, we assess the performance of split NLC via numerical simulations, and characterize performance in terms of achievable signal-to-noise ratio (SNR) and mutual information (MI).  Further, we confirm these results theoretically.


\section{Model of Transmission Performance Using Digital Nonlinearity Compensation} \label{sec:theory}

To model the effect of digital NLC we used a coherent Gaussian noise (GN) model of nonlinear interference \cite{Poggiolini2012} including the effect of signal-ASE (amplified spontaneous emission) noise interactions.  The model treats the field propagating in the fiber as a summation of signal and noise fields, incorporating the signal-ASE noise interaction as a form of cross channel interference. 
{Similar to \cite{Ellis2015}} the symbol SNR (signal-to-noise ratio) at the receiver is approximated as
\begin{equation}
\text{SNR} \approx \frac{P}{N_S P_{ASE} + N_S^{1+\varepsilon_{ss}} \eta_{ss} P^3 + 3 \xi \eta_{sn} P^2 P_{ASE}},
\label{eqn:SNR}
\end{equation}
where $P$ is the signal power, $N_S$ is the number of spans, $P_{ASE}$ is the ASE noise power in the signal bandwidth from a single span amplifier, $\eta_{ss}$ is a single span nonlinear interference factor for the self channel interference (i.e., signal-signal interactions), $\eta_{sn}$ is a single span nonlinear interference factor for signal-squared ASE noise interference, $\varepsilon_{ss}$ is the coherence factor for self channel interference and $\xi$ is a factor depending on the number of spans and the method used for digital NLC. The nonlinear interference terms with ASE noise squared and cubed have been neglected as insignificant.  In order to analytically calculate the SNR when applying different nonlinear equalization methods, the $\xi$ parameter must be computed for DPC, DBP and split NLC ($\xi_{DPC}$, $\xi_{DBP}$ and $\xi_{SC}$, respectively).

Following the method in the Appendix, it is found that the difference in SNR at optimum signal launch power when applying split NLC versus DBP is given by
\begin{equation}
\Delta \text{SNR}=\sqrt{\frac{\xi_{DBP}}{\xi_{SC}}},
\label{eq:snrgain1}
\end{equation}
where $\xi_{DBP}$ and $\xi_{SC}$ are given in the appendix by Eqs.~\eqref{eqn:xi_DBP} and \eqref{eqn:xi_SC}. Choosing a 50\% transmitter:receiver split ratio for NLC, and for a large number of spans, \eqref{eq:snrgain1} becomes
\begin{equation}
\lim_{N_S \to \infty} \Delta \text{SNR}=\sqrt{2^{1+\varepsilon_{sn}}}.
\label{eq:snrgain2}
\end{equation}
For this work, $\varepsilon_{sn}=0.134$. For $\varepsilon_{sn}$ varying from 0 to 0.3 (a conservatively high value) $\Delta \text{SNR}$ varies between 1.5~dB and 1.95~dB.  In the large bandwidth limit, $\varepsilon_{sn}$ tends to zero.

\section{Numerical Simulations}
Consider the point-to-point transmission link shown in Fig.~\ref{fig:config}, consisting of an idealized optical transmitter and coherent receiver separated by $N_S$ spans of standard single mode fiber (SSMF), followed by erbium doped fiber amplifiers (EDFA).  The simulation parameters are shown in Table~\ref{tab:parameters}.

The transmitted signal was single channel 50~GBd polarization division multiplexed 256-\textit{ary} quadrature amplitude modulation (PDM-256QAM).  This format was chosen as it is sufficiently high cardinality to demonstrate increases in MI when using NLC for all transmission distances considered.  The signal was sampled at 4~samples/symbol (to take account of the signal broadening due to fiber nonlinearity) and shaped using a root-raised cosine (RRC) filter.  Where DPC was considered, it was applied at this point using the split step Fourier method (SSFM) to solve the Manakov equation \cite[Eq.~(12)]{ManakovEq}.

The optical fiber span was again modeled by solving the Manakov equation using the SSFM.  Each fiber span was followed by an EDFA which applied gain which exactly compensated the previous span loss.

Where required, the receiver DSP applied either frequency domain chromatic dispersion compensation (linear case), or DBP.  Subsequently, a matched RRC filter was applied to the signal, and the signal was downsampled to 1~sample/symbol.  To mitigate any residual phase rotation due to uncompensated nonlinear interference, carrier phase recovery was performed as described in \cite{Alvarado2015}.  Finally, the SNR was estimated over $2^{17}$ symbols by comparing the transmitted and received symbols as in \cite{Alvarado2015}.  For SSFM simulations, the MI was computed using Monte Carlo integration and is included as a figure of merit to provide insight into the gains in throughput possible when employing digital NLC.  Note that the analytical MI results are obtained using numerical integration.


\begin{table}
	\centering
	\caption{Summary of system parameters used in fiber simulation}

	\begin{tabular}{ l | r | l }
\hline
	Parameter		 				& Value & Units \\
\hline 

\hline
		Fiber attenuation			& $0.2$&$\tnr{dB/km}$ 				\\
		Dispersion parameter			& $17$&$\tnr{ps}/(\tnr{nm}\cdot\tnr{km})$	\\
		Fiber nonlinear coefficient		& $1.2$&$1/(\tnr{W}\cdot\tnr{km})$ 			\\
		Span length				& $100$&$\tnr{km}$					\\
		Simulation step size				& $100$&$\tnr{m}$					\\
		NLC step size				& $100$&$\tnr{m}$					\\
		Symbol rate				& $50$&$\tnr{GBd}$				\\
		EDFA noise figure 			& $5$&$\tnr{dB}$					\\
		Pulse shape				& RRC, $1\%$ & rolloff				\\
\hline

\hline
	\end{tabular}
	\label{tab:parameters}
\end{table}%

%
%
%

\section{Results and Discussion}
\pgfplotsset{compat = 1.3}
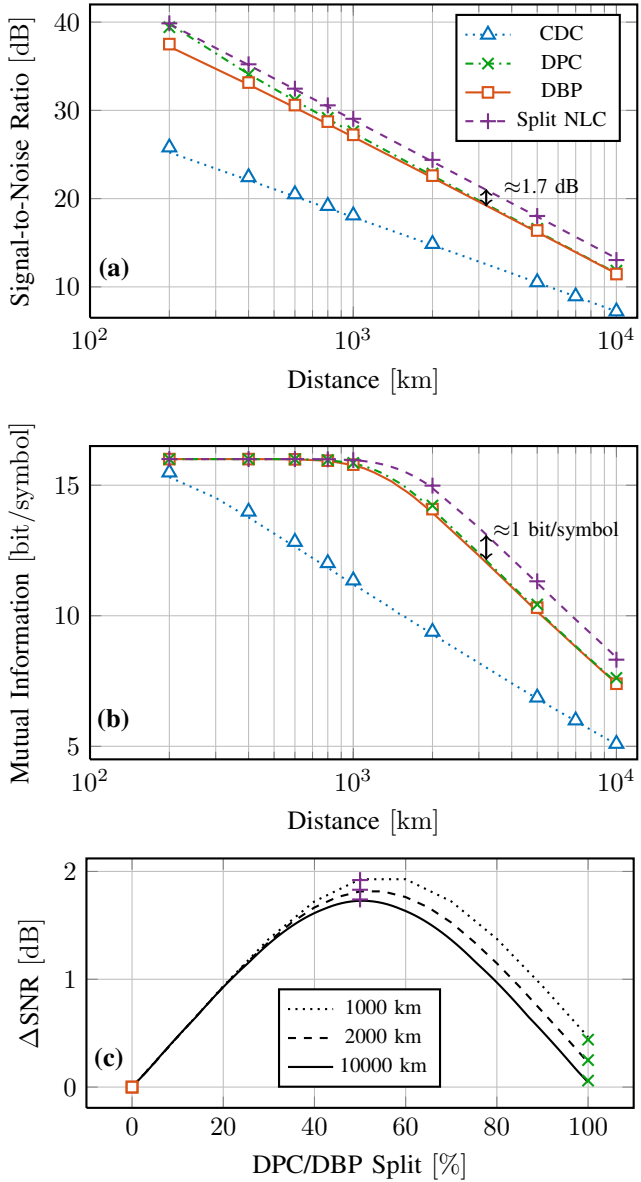
\begin{figure}[!tb]
\centering
\begin{tabular}{c}
\begin{tikzpicture}[every pin/.style={fill=white}]
	\begin{semilogxaxis}[
	name=masterplot,
	height=5.75cm,width=\columnwidth,
	xlabel={Distance},
	y unit={dB},
	ylabel={Signal-to-Noise Ratio},
	x unit={km},
	xmin=9.99e1,
	xmax=1.2e4,
	ymin=6.5,
	ymax=42, 
	grid=both,
	legend entries={\footnotesize CDC,\footnotesize DPC,\footnotesize DBP,\footnotesize Split NLC},
	legend pos=north east,
	thick, 
	]

	\definecolor{s1}{RGB}{0,  113.9850,  188.9550}
	\definecolor{s2}{RGB}{216.7500,   82.8750,   24.9900}
	\definecolor{s3}{RGB}{10,   160,   20}
	\definecolor{s4}{RGB}{125.9700,   46.9200,  141.7800}

	\addlegendimage{style=dotted,mark=triangle*,mark size=3,draw=s1,mark options={fill=white,style=solid}}
	\addlegendimage{style=dashdotted,mark=x,mark size=3,draw=s3,mark options={solid}}
	\addlegendimage{style=solid,mark=square*,mark size=2,draw=s2,mark options={fill=white,style=solid}}
	\addlegendimage{style=dashed,mark=+,mark size=3,draw=s4,mark options={solid}}	

	\pgfplotstableread{Results/cdc_sim_snr.txt}
	\datatable%
	\addplot +[only marks,mark=triangle*,mark size=3,draw=s1,mark options={fill=white,style=solid}] table [x index=0, y index=1] {\datatable};

	\pgfplotstableread{Results/dpc_sim_snr.txt}
	\datatable%
	\addplot +[only marks,mark=x,mark size=3,draw=s3] table [x index=0, y index=1] {\datatable};	

	\pgfplotstableread{Results/dbp_sim_snr.txt}
	\datatable%
	\addplot +[only marks,mark=square*,mark size=2,draw=s2,mark options={fill=white,style=solid}] table [x index=0, y index=1] {\datatable};	
	
	\pgfplotstableread{Results/split_sim_snr.txt}
	\datatable%
	\addplot +[only marks,mark=+,mark size=3,draw=s4] table [x index=0, y index=1] {\datatable};		

	\pgfplotstableread{Results/cdc_ana_snr.txt}
	\datatable%
	\addplot +[mark=none,draw=s1,style=dotted] table [x index=0, y index=1] {\datatable};

	\pgfplotstableread{Results/dpc_ana_snr.txt}
	\datatable%
	\addplot +[mark=none,draw=s3,style=dashdotted] table [x index=0, y index=1] {\datatable};

	\pgfplotstableread{Results/dbp_ana_snr.txt}
	\datatable%
	\addplot +[mark=none,draw=s2,style=solid] table [x index=0, y index=1] {\datatable};
	
	\pgfplotstableread{Results/split_ana_snr.txt}
	\datatable%
	\addplot +[mark=none,draw=s4,style=dashed] table [x index=0, y index=1] {\datatable};

	


	\node[anchor=south] (source) at (axis cs:5200,19.50){\footnotesize{$\approx$1.7~dB}};	
	
      \node[anchor=south] (source) at (axis cs:3200,21.10){};
       \node (destination) at (axis cs:3200,18.00){};
       \draw[<->](source)--(destination);	


\node[black,left] at (axis cs:154,12){\normalsize{\textbf{(a)}}};

	\end{semilogxaxis}
\end{tikzpicture}
\\
\begin{tikzpicture}[every pin/.style={fill=white}]
	\begin{semilogxaxis}[
	height=5.750cm,width=\columnwidth,
	xlabel={Distance},
	y unit={bit/symbol},
	ylabel={Mutual Information},
	x unit={km},
	xmin=9.99e1,
	xmax=1.2e4,
	ymin=4.5,
	ymax=16.5,
	grid=both,
	thick, 
	]
	
	\definecolor{s1}{RGB}{0,  113.9850,  188.9550}
	\definecolor{s2}{RGB}{216.7500,   82.8750,   24.9900}
	\definecolor{s3}{RGB}{10,   160,   20}
	\definecolor{s4}{RGB}{125.9700,   46.9200,  141.7800}

	\addlegendimage{style=dotted,mark=triangle*,mark size=3,draw=s1,mark options={fill=white,style=solid}}
	\addlegendimage{style=solid,mark=square*,mark size=2,draw=s2,mark options={fill=white,style=solid}}
	\addlegendimage{style=dashdotted,mark=x,mark size=3,draw=s3,mark options={solid}}
	\addlegendimage{style=dashed,mark=+,mark size=3,draw=s4,mark options={solid}}

	\pgfplotstableread{Results/cdc_ana_mi.txt}
	\datatable%
	\addplot +[mark=none,draw=s1,style=dotted] table [x index=0, y index=1] {\datatable};
	
	\pgfplotstableread{Results/dbp_ana_mi.txt}
	\datatable%
	\addplot +[mark=none,draw=s2,style=solid] table [x index=0, y index=1] {\datatable};

	\pgfplotstableread{Results/dpc_ana_mi.txt}
	\datatable%
	\addplot +[mark=none,draw=s3,style=dashdotted] table [x index=0, y index=1] {\datatable};
	
	\pgfplotstableread{Results/split_ana_mi.txt}
	\datatable%
	\addplot +[mark=none,draw=s4,style=dashed] table [x index=0, y index=1] {\datatable};	

	\pgfplotstableread{Results/cdc_sim_mi.txt}
	\datatable%
	\addplot +[only marks,mark=triangle*,mark size=3,draw=s1,mark options={fill=white}] table [x index=0, y index=1] {\datatable};
	
	\pgfplotstableread{Results/dbp_sim_mi.txt}
	\datatable%
	\addplot +[only marks,mark=square*,mark size=2,draw=s2,mark options={fill=white,style=solid}] table [x index=0, y index=1] {\datatable};	

	\pgfplotstableread{Results/dpc_sim_mi.txt}
	\datatable%
	\addplot +[only marks,mark=x,mark size=3,draw=s3] table [x index=0, y index=1] {\datatable};	
	
	\pgfplotstableread{Results/split_sim_mi.txt}
	\datatable%
	\addplot +[only marks,mark=+,mark size=3,draw=s4] table [x index=0, y index=1] {\datatable};	

	

	\node[anchor=south] (source) at (axis cs:5900,12.50){\footnotesize $\approx$1 bit/symbol};	
	
      \node[anchor=south] (source) at (axis cs:3200,13.10){};
       \node (destination) at (axis cs:3200,11.75){};
       \draw[<->](source)--(destination);	
	

\node[black,left] at (axis cs:154,6){\normalsize{\textbf{(b)}}};
	\end{semilogxaxis}
\end{tikzpicture}
\\
\begin{tikzpicture}[every pin/.style={fill=white}]
\begin{axis}[
	height=4.9cm,width=\columnwidth,
	xlabel={DPC/DBP Split},
	y unit={dB},
	ylabel={$\Delta$SNR},
	x unit={\%},
	grid=both,
	legend entries={\footnotesize 1000~km, \footnotesize 2000~km, \footnotesize 10000~km},
	legend style={at={(0.35,0.28)},anchor=west},
	thick, 
]
	
	\definecolor{s1}{RGB}{0,  113.9850,  188.9550}
	\definecolor{s2}{RGB}{216.7500,   82.8750,   24.9900}
	\definecolor{s3}{RGB}{10,   160,   20} 
	\definecolor{s4}{RGB}{125.9700,   46.9200,  141.7800}	
	
	\pgfplotstableread{Results/split1000km.txt}
	\datatable%
	\addplot +[mark=none,draw=black,style=dotted] table [x index=0, y index=1] {\datatable};

	\pgfplotstableread{Results/split2000km.txt}
	\datatable%
	\addplot +[mark=none,draw=black,style=dashed] table [x index=0, y index=1] {\datatable};
	
	\pgfplotstableread{Results/split10000km.txt}
	\datatable%
	\addplot +[mark=none,draw=black,style=solid] table [x index=0, y index=1] {\datatable};


	\addplot[only marks,mark=square*,mark size=2,draw=s2,mark options={fill=white,style=solid}] coordinates {(0,0)};
	\addplot[only marks,mark=x,mark size=3,draw=s3] coordinates {(100,0.44)};
	\addplot[only marks,mark=+,mark size=3,draw=s4] coordinates {(50,1.92)};

	\addplot[only marks,mark=square*,mark size=2,draw=s2,mark options={fill=white,style=solid}] coordinates {(0,0)};
	\addplot[only marks,mark=x,mark size=3,draw=s3] coordinates {(100,0.25)};
	\addplot[only marks,mark=+,mark size=3,draw=s4] coordinates {(50,1.83)};
	
	\addplot[only marks,mark=square*,mark size=2,draw=s2,mark options={fill=white,style=solid}] coordinates {(0,0)};
	\addplot[only marks,mark=x,mark size=3,draw=s3] coordinates {(100,0.06)};
	\addplot[only marks,mark=+,mark size=3,draw=s4] coordinates {(50,1.74)};	

\node[black,left] at (axis cs:0.3,0.26){\normalsize{\textbf{(c)}}};
\end{axis}		
\end{tikzpicture}

\end{tabular}
\caption{Analytical (curves) and simulated (markers) transmission performance when applying different fiber nonlinearity compensation methods.  (a) SNR (over  signal bandwidth) vs. transmission distance, (b) MI vs. transmission distance, and (c) SNR gain over DBP when varying NLC split ratio.}
\label{fig:results}
\end{figure}
The analytical expression for SNR, \eqref{eqn:SNR}, was evaluated using the methods outlined in the Appendix for calculating $\xi$ for both linear signal equalization (CDC) and for each NLC technique: DPC, DBP and split NLC. In simulation, transmission distances were considered between 200 and 10000~km (2-100~spans), with the signal launch power varied in 1~dB steps.  For each transmission distance, the MI and SNR were determined at the optimum launch power.  Fig.~\ref{fig:results}(a) shows how the maximum achievable SNR varies with transmission distance when applying different digital NLC techniques.  A 50\% split ratio is used for split NLC.  It should be noted that there is excellent agreement between the analytical expressions and the SSFM simulations, with an SNR estimation accuracy better than 0.5~dB for short distances, and better than 0.2~dB for distances above 1000~km, where the GN model is known to have greater accuracy due to the high accumulated chromatic dispersion.  An SNR improvement for split NLC over both DBP and DPC at all distances is also observed, as predicted by \eqref{eq:snrgain2}.

Fig.~\ref{fig:results}(b) shows how the MI of the received signal degrades with distance when applying different digital techniques for equalization.  The modulation format considered (DP-256QAM) encodes a maximum 16 bits of information.  The advantage of digital NLC is clear as, even at short distances, this maximum MI cannot be achieved without either DPC, DBP or split NLC.  In longer reach scenarios ($>$1000~km) there is a clear gain in MI when using split NLC compared with DPC or DBP, and this gain saturates for distances greater than approximately 2000~km to be approximately 1~bit/symbol.

The results in Fig.~\ref{fig:results}(c) show the SNR gain that can be achieved by dividing the NLC between transmitter and receiver with different ratios.  Note that the gain of DPC over DBP rapidly diminishes with transmission distance.  Further, it can be seen that a 50\% NLC split ratio is optimum for all transmission distances.

It should be noted that the SSFM simulations and the theoretical analysis represent a somewhat idealized model of an optical fiber transmission system.  For example, polarization mode dispersion is known to negatively impact on the performance of digital NLC, and yet is not considered in this model.  Therefore, these results should be interpreted as an optimistic estimation of performance using digital NLC.  Nevertheless, this work demonstrates that the current arrangements of digital NLC (DPC or DBP) can be substantially improved.

\section{conclusions}
We used a closed form approximation for the accumulation of the signal-ASE interaction over multiple spans in order to analyse the potential SNR gain when dividing NLC between transmitter and receiver.  The optimum launch power, and hence SNR gain, when using split NLC will increase by 1.5~dB with respect to both DPC and DBP in the limit of long distance, high bandwidth transmission.  Split NLC is shown, both theoretically and by numerical simulation, to globally outperform both DPC and DBP for all transmission distances.  There is scope to use this SNR gain to reduce the complexity of NLC by dividing the DSP between transmitter and receiver; a subject for further investigation.

\appendix
{The following is a derivation of analytical expressions for $\xi$ in the case of both linear CDC at the receiver, and nonlinear compensation using pre-, post- or split-NLC.} The nonlinear interference factors, $\eta_{ss}$ and $\eta_{sn}$, were calculated using numerical integration of the GN model reference equation \cite[Eq. (1)]{Poggiolini2012}.
%
%
Note $\eta_{sn}\approx\eta_{ss}$ but is more accurately given by numerical integration of \cite[Eq. (7)]{Ives2014b} where the spectral shape, $g(\dot f_1 + \dot f_2 + f)$, is replaced by {unity} to represent the uniformity of the ASE spectrum. Coherence factors $\varepsilon_{ss}$ and $\varepsilon_{sn}$ were calculated by obtaining the nonlinear interference factors for 100 spans by numerical integration and using
\begin{equation}
\varepsilon = \frac{\log{\left(\frac{\eta_{100}}{\eta_1}\right)}}{\log{\left(100\right)}}-1
\label{eqn:epsilon}
\end{equation}
where $\eta_1$ is the single span nonlinear interference factor and $\eta_{100}$ is the nonlinear interference factor for 100 spans. In each case, $\eta$ is substituted by $\eta_{ss}$ or $\eta_{sn}$, as appropriate. Note that the purpose of $\varepsilon$ is to change the coherence of the interference terms, altering the accumulation of the nonlinear interference with number of spans, and that $0\le\varepsilon\le1$.

The effect of NLC on the effective received SNR is modeled by assuming that NLC effectively subtracts in power the nonlinear interference generated by the forward propagating field. 
This simplification is customary in the literature and can be seen as the result of two assumptions: i) a perturbative first-order approximation, and ii) uncorrelation of all the optical fields involved in the SNR calculation. 
As shown in \cite{Secondini2014}, DBP generates a first-order field, identical, but with opposite sign, to the forward-propagated field, provided that the linearly-propagated field (zeroth-order solution) along the fiber link is the same, hence the cancellation.     
However, due to the noise accumulation over the link, there is a mismatch between the linearly forward-propagated field and the backward-propagated field (or the precompensated field).  
As a result, residual signal-ASE interaction terms are still present after the application of either DBP or DPC, representing one of the main performance limitations \cite{Gao2012}.  

Further assuming a weak nonlinear interaction between ASE noise contributions along the link\footnote{Indeed, the perturbative approximation does not allow the additivity of nonlinear terms  arising from two or more optical fields adding together, even to the first order.}, the calculation of the signal-ASE interaction terms can be performed by considering each ASE noise contribution as separately interacting with the signal in each span.

The nonlinear interference scaling coefficient, $\xi$ accounts for the noise generated due to this signal-ASE interaction.  In the case of linear chromatic dispersion compensation (CDC), each contribution of ASE noise interacts with the signal from the span following its addition, up until the end of the link. In the configuration analysed herein (Fig.~\ref{fig:config}), the first ASE noise contribution interacts with the signal in the second span.  Likewise, in the case of DPC, signal-ASE noise interference accumulates from the second span onwards, since the first noise source follows the first span. Thus, for the CDC and DPC scenarios, 
\begin{equation}
\xi_{DPC} = \xi_{CDC} = \sum_{k=1}^{N_S-1} k^{1+\varepsilon_{sn}}.
\label{eqn:xi_DPC}
\end{equation}

For DBP, noise from the last amplifier will be backpropagated as if it were signal for $N_S$ spans. 
The noise from the penultimate amplifier will have interacted with the signal for one span but will be backpropagated as if it were signal over all $N_S$ spans. 
Thus the signal noise interaction over the final span will be correctly compensated, leading to $N_S-1$ spans of excess nonlinear interference. 
Thus the total signal-ASE noise interference is given by the following sum over all spans
\begin{equation}
\xi_{DBP} = \sum_{k=1}^{N_S} k^{1+\varepsilon_{sn}}.
\label{eqn:xi_DBP}
\end{equation}

If the NLC is split between $N_{S1}$ spans of DPC and $N_{S2}$ spans of DBP such that the total number of spans is $N_S = N_{S1} + N_{S2}$, then $\xi_{SC}$ is given by
\begin{equation}
\xi_{SC} = \sum_{k=1}^{N_{S1}-1} k^{1+\varepsilon_{sn}} + \sum_{k=1}^{N_{S2}} k^{1+\varepsilon_{sn}}.
\label{eqn:xi_SC}
\end{equation}
The advantage of splitting the compensation arises since $\xi_{DBP}$ and $\xi_{DPC}$ increase superlinearly with the number of spans. 
$\xi_{SC}$ is minimized for $N_{S1} = \left \lceil \frac{N_S}{2} \right \rceil$ and $N_{S2} = \left \lfloor \frac{N_S}{2} \right \rfloor$.

The SNR gain due to the split NLC can be quantified using an approximated closed-form expression for the summation, in each of \eqref{eqn:xi_DPC} and \eqref{eqn:xi_DBP}. Using Faulhaber's formula \cite[Eq.~CE~332, pg.~1]{Gradshteyn2000}, such a summation can be expressed as
\begin{multline}
\label{eqn:closedform}
\sum_{k=1}^{N_S}k^{1+\varepsilon_{sn}}=\frac{N_{S}^{2+\varepsilon_{sn}}}{2+\varepsilon_{sn}}+\frac{N_{S}^{1+\varepsilon_{sn}}}{2}+\\
\frac{1}{2}\binom{1+\varepsilon_{sn}}{1}B_2 N_S^{\varepsilon_{sn}}+\frac{1}{4}\binom{1+\varepsilon_{sn}}{3}B_4 N_S^{\varepsilon_{sn}-2}+...
\end{multline}
where the coefficients $B_n$ are known as the Bernoulli numbers. 
A sufficiently accurate closed-form for $N_{S}>1$ can be derived by truncating \eqref{eqn:closedform} to the first 2 terms. These terms rapidly dominate the higher order terms as $N_S$ increases, particularly considering that $B_2=1/6$ and $B_4=-1/30$.
The SNR gain, $\Delta \text{SNR}$, for split compensation over DBP can be defined as the ratio between the SNRs achieved by each compensation technique at optimum launch power. All NLC techniques remove the cubic terms in \eqref{eqn:SNR}. Thus, considering that maximizing the SNR leads to the optimum launch power given by
\begin{equation}
P_{opt}=\frac{1}{2P_{ASE}\sqrt{N_S\eta_{sn}\xi}}
\end{equation} 
and that $\text{SNR}\propto{}P$ at the optimum power since the overall ASE noise power is equal to the signal-ASE interaction power, the change in SNR is given by \eqref{eq:snrgain1}.
%
%
%
%
Substituting the first two terms from the approximation \eqref{eqn:closedform} into \eqref{eq:snrgain1}, and choosing $N_{S1} = \left \lceil \frac{N_S}{2} \right \rceil$ when calculating $\xi_{SC}$, for an asymptotically large number of spans, we obtain \eqref{eq:snrgain2}.

\section*{Acknowledgment}
The authors wish to thank Prof.~A.~Ellis for useful discussions and comments on an earlier draft of this Letter.

\ifCLASSOPTIONcaptionsoff
  \newpage
\fi


\begin{thebibliography}{12}
\bibitem{Liu2013}
X.~Liu, A.~R.~Chraplyvy, P.~J.~Winzer, R.~W.~Tkach, and S.~Chandrasekhar, ``Phase-conjugated twin waves for communication beyond the Kerr nonlinearity limit,'' \textit{Nat. Photon}, vol. 7, no. 7, pp. 560--568, Apr. 2013.


\bibitem{Ellis2015_OPC}
A.~D.~Ellis, M.~E.~McCarthy, M.~A.~Z.~Al-Khateeb, and S.~Sygletos, ``Capacity limits of systems employing multiple optical phase conjugators,'' \textit{Opt. Express}, vol.~23, no.~16, pp.~20381--20393, Aug. 2015.

\bibitem{Temprana2015}
E.~Temprana, \textit{et al.}, ``Two-fold transmission reach enhancement enabled by transmitter-side digital backpropagation and optical frequency comb-derived information carriers,'' \textit{Opt. Express}, vol. 23, no. 16, pp.~20774--20783, Aug. 2015.

\bibitem{Roberts2006}
K.~Roberts, L.~Chuandong, L.~Strawczynski, M.~O'Sullivan, I.~Hardcastle, ``Electronic precompensation of optical nonlinearity,'' \textit{IEEE Photon. Technol. Lett.}, vol.~18, no.~2, pp.~403--405, Jan. 2006.

\bibitem{Maher2015}
R. Maher, \textit{et al.}, ``Reach Enhancement of 100\% for a DP-64QAM Super-Channel using MC-DBP,'' \textit{in Proc.} OFC Conf., Mar 2015, pp.~1--3, paper Th4D.5.

\bibitem{Lowery2007}
A.~J.~Lowery, ``Fiber nonlinearity pre- and post-compensation for long-haul optical links using OFDM,'' \textit{Opt. Express}, vol.~15, no.~20, pp.~12965--12970, Oct. 2007.

\bibitem{Lavery2015}
D.~Lavery, \textit{et al.}, ``Low Complexity Multichannel Nonlinear Predistortion for Passive Optical Networks,'' \textit{in Proc.} Signal Processing in Photonic Communications Conference, May 2015, paper SpS2C.5.

\bibitem{Gao2013}
Y.~Gao, \textit{et al.}, ``Reducing the Complexity of Perturbation based Nonlinearity Pre-compensation Using Symmetric EDC and Pulse Shaping,'' \textit{Opt. Express}, vol. 22, no. 2, pp.~1209--1219, Jan. 2014.

\bibitem{Secondini2014a}
M.~Secondini and E.~Forestieri, ``On XPM Mitigation in WDM Fiber-Optic Systems'' \textit{IEEE Photon. Technol. Lett.}, vol.~26, no.~22, pp.~2252--2255, Nov. 2014.

\bibitem{Liga2015}
G.~Liga, \textit{et al.} ``Optimum Detection in Presence of Nonlinear Distortions with Memory,'' \textit{in Proc.} ECOC, Sep. 2015, pp.~1--3, Paper P.4.13.

\bibitem{Poggiolini2012}
P.~Poggiolini, ``{The GN Model of Non-Linear Propagation in Uncompensated Coherent Optical Systems},'' \textit{J. Lightwave Technol.}, {vol.~30}, {no.~24}, pp.~3857--3879, Dec. 2014.

\bibitem{Ellis2015}
A.~D.~Ellis, \textit{et al.}, ``The impact of phase conjugation on the nonlinear-Shannon limit: The difference between optical and electrical phase conjugation,''. \textit{in Proc.} IEEE Summer Topical Meeting on Nonlinear Optical Signal Processing, Nassau, BS, vol. 2, pp. 209--210, Jul. 2015.

\bibitem{ManakovEq}
D.~Marcuse, C.~R.~Menyuk, and P.~K.~A.~Wai, ``Application of the Manakov-PMD equation to studies of signal propagation in optical fibers with randomly varying birefringence,'' \textit{J. Lightwave Technol.}, vol.~15, no.~9, pp.~1735--1746, Sep. 1997.

\bibitem{Alvarado2015}
A.~Alvarado, E.~Agrell, D.~Lavery, R.~Maher, and P.~Bayvel, ``Replacing the Soft-decision FEC Limit Paradigm in the Design of Optical Communication Systems,'' \textit{J. Lightwave Technol.}, vol.~33, no.~20, pp.~4338--4352, Oct. 2015.

\bibitem{Ives2014b}
D.~J.~Ives, P.~Bayvel, and S.~J.~Savory, ``{Adapting Transmitter
  Power and Modulation Format to Improve Optical Network Performance Utilizing the Gaussian Noise Model of Nonlinear Impairments},'' \textit{J. Lightwave Technol.}, {vol.~32}, {no.~21}, pp.~3485--3494, Nov. 2014.

\bibitem{Secondini2014}
M.~Secondini, E.~Forestieri, and G.~Prati, ``Achievable Information Rate in Nonlinear WDM Fiber-Optic Systems With Arbitrary Modulation Formats and Dispersion Maps'' \textit{J. Lightwave Technol.}, vol.~31, no.~23, pp.~3839--3852, Dec.~2013.

\bibitem{Gao2012}
G.~Gao, \textit{et al.},``Influence of PMD on fiber nonlinearity compensation using digital back propagation'', \textit{Opt. Express}, {vol.~20}, {no.~2}, pp.~{14406--14418}, Jun.~2012.

\bibitem{Gradshteyn2000}
I.~S.~Gradshteyn and I.~M.~Ryzhik, ``Table of Integrals, Series, and Products'', 6th Edition, Academic Press, 2000.

\end{thebibliography}
\end{document}